# Low Radon Concentration Measurement with a Radon-Dissolved Liquid Scintillator Pilot Detector[*]

Qian-Yun Li,[1] Ming-Xuan Li,[1] Ren-Ming-Jie Li,[1] Shin-Ted Lin,[1,†] Shu-Kui Liu,[1,‡]
Chang-Jian Tang,[1] Hao-Yang Xing,[1] Jing-Jun Zhu,[1] Qian Yue,[3] and Li-Tao Yang[3]

[1] College of Physics, Sichuan University, Chengdu 610065, China
[2] Institute of Nuclear Science and Technology, Sichuan University, Chengdu 610064, China
[3] Key Laboratory of Particle and Radiation Imaging (Ministry of Education) and
Department of Engineering Physics, Tsinghua University, Beijing 100084, China

We construct a high sensitivity radon pilot detector using liquid scientillator dissolved radon for the CDEX rare-event searches program. The CDEX-10 project employs a germanium detector array immersed in a massive liquid nitrogen tank. However, radon emanated from the surface of the tank can contribute background. As a result, radon contamination in the liquid nitrogen tank must be regulated and monitored, which necessitates the use of a low-level Rn measuring device. The radon-dissolved liquid scientillator (LS) detectors utilizes cascade decayselections, yielding a unique signature. All background events occurring in the LS, with regard to the system and co-efficiency of Rn absorption, have been studied. Background activity is mearsured to be $(59.8 \pm 18.4)\,\mu Bq$ per $300\,mL$ LS. Meanwhile, it shows that radon concentration absorbed by LS will be saturated and is linearly related to radon concentration in nitrogen gas. Furthermore, the co-efficiency factor can be optimized by lowering temperature or raising pressure. Eventually, the detection limit for $^{222}Rn$ in nitrogen gas is observed to be $9.6\,mBq/m^3$.

## I. INTRODUCTION

Because the interaction cross-section between WIMPs and normal matter is extremely small [1], experiments geared toward dark matter search have to be performed in underground laboratories with extremely low background environments [2]. The China Dark Matter Experiment (CDEX) [3] is the first underground investigation on dark matter search in China. It will be performed at the China JinPing Underground Laboratory (CJPL) [4], which is covered by mountains that are $2400\,m$ of rock layer, to realize extremely low radioactivity [5]. Currently, CDEX is progressing toward CDEX-10, which is a project wherein a p-type point contact germanium ($p$PCGe) detector array has been deployed in a liquid nitrogen tank that exhibits a diameter and height of $13\,m$ [6]. The vast liquid nitrogen acts as a cooling medium for germanium detector arrays as well as a passive shielding material against ambient radioactivity.

However, $^{222}Rn$ and its radioactive daughters, especially long-living $^{210}Pb$, are a significant source of background events that occur in CDEX[7, 8]. What's worse, radon concentration in environment of CJPL is about 10 times in nature [9]. The radon content in and around the detector has to be decreased to a tolerable level; this requires precise low-level Rn measurements.

Proportional counting after radon collection from the bulk sample is the most sensitive method for detecting alpha decays from $^{222}Rn$ and its daughters [10] However, this method is unsuitable for continuous monitoring.

A well-known method for $^{222}Rn$ detection in air is electrostatic collection of the $^{222}Rn$ daughter $^{218}Po$ on an alpha detector, which is followed by the measurement of the subsequent alpha decays of $^{218}Po$ and $^{214}Po$ [11,12]. This method is utilized by many rare event experimental collaborations [13−16]. The detector used in Super-Kamiokande neutrino experiment exhibits an active volume of 68.7 L, with a detection limit of $1.4\,mBq/m^3$ $^{222}Rn$ in air [13]. Later, the Super-Kamiokande detector was improved by increasing its active volume and augmenting its voltage to get a detection limit of $0.7\,mBq/m^3$ [15]. However, these steps were inadequate for the optimization of this method. Accordingly, the BOREXINO detector focuses on background control and achieved a detection limit of $0.07\,mBq/m^3$ [14].

With radon detection techniques developed, radon concentration background is measured to be $48.0\,\mu Bq/kg$ in liquid xenon of $62\,kg$ [17] which is similar to our condition. Furthermore, with some other later results[18−21], it shows that radon concentration background is approximately inversely proportional to two-third power volume.So that a detection limit of $1\,mBq/m^3$ for $^{222}Rn$ in liquid nitrogen of $1700\,m^3$ is considered to be apt for CDEX-10. To achieve this goal,, a prototype measurement system that utilizes cascade decay selection and liquid scintillator (LS) detectors has been developed. There are 3 major reasons for adopting this route: First, LS exhibits a significantly high purity with extremely low U/Th contamination; this is in contrast to most other materials that exhibit U/Th contamination, which leads to intrinsic Rn contamination. Secondly, based on past experience, radon solubility of LS is rather high, which is useful for lowering the detection limit. The last and the most important reason is that cascade decay selection produces highly efficient and precise results because it considers the uniqueness

[*] The National Key Research and Development Program of China (Contract No. 2023YFA1607, 2017YFA0402203), the National Natural Science Foundation of China (Contracts No. 11975159, No. 11975162), and the Sichuan Provincial Natural Science Foundation (No. 2022NSFSC1825) provided support for this work. The authors express their gratitude to Nanhua University for supplying the radon sources.
[†] Corresponding author, ShinTed Lin: stlin@scu.edu.cn
[‡] Corresponding author, Shukui Liu: liusk@scu.edu.cn

of certain decay half-lives [22,23].

## II. EXPERIMENTAL SETUP AND PROCESS

### A. Experimental setup

For measuring the excessively low radon concentration, the materials used to develop the measurement system must exhibit have minimal radon emanation. In general, the electropolished stainless could meet these requirements [24, 25]; therefore, the system used in this study except detector is entirely made up of electropolished stainless steel, which has been subjected to pickling and passivation. As shown in Fig. 1, the experimental system comprises four major parts: storage tank, mixing tank, exhaust gas treatment system, and detector.

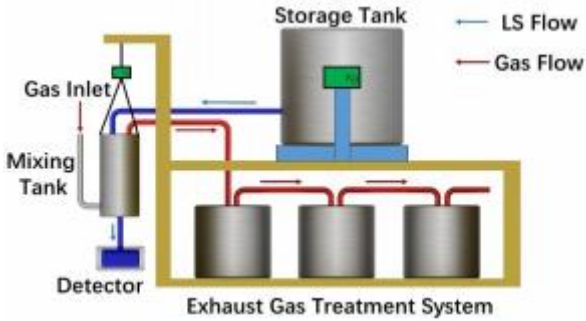

Fig. 1. Schematic diagram of the experimental system.

The storage tank with a total volume of $30L$ is used to safely store LS (EJ-309) and situated on an electronic scale to monitor its mass. It is crucial to store LS in the storage tank to avoid exposure to ambient air during during its use and transfer of LS. In order to reduce the impact of weighing on the mixing tank, we utilize polytetrafluoroethylene (PTFE) piping, which is characterized by its low radon emanation and high air tightness. The mixing tank with a total volume of $1.6L$ is suspended above an electronic scale. Its design is slender so that carrier gas can have maximum contact with LS, which could enhance the absorption efficiency. Due to its volatility, flammability and toxicity, the carrier gas exit is connected to an appropriate exhaust gas treatment system, consisting of three interconnected tanks filled with LS EJ-309, molecular sieves, and ethanol in sequence.

Fig. 2 shows detector design. The detector with a total volume of $0.3L$ is made up of standard JGS1, high-purity quartz, and a Teflon layer that is wrapped around the cylinder to enhance reflectivity for scintillation light. The photomultiplier tube(PMT) from Beijing HAMAMATSU(type CR173) located at the base of the cylinder is coupled by silicone grease. The basic and direct Data AcQuisition (DAQ) system is illustrated in Fig. 3. Our DAQ system is based on NI PXIe-5170R oscilloscope manufactured by National Instrument.Its internal controller is comprised of a reprogrammable FPGA, which enables programming of the trigger type as a logical

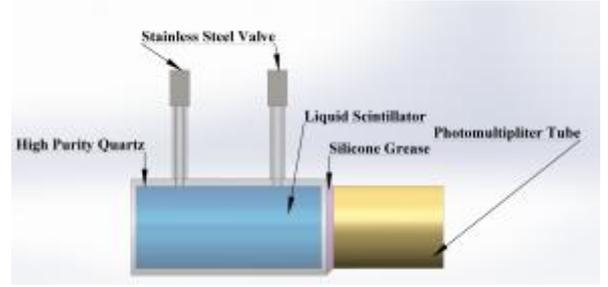

Fig. 2. Schematic diagram of the detector.

OR gate; this ensures that the PMT signal can directly transits to the oscilloscope. The PXIe-5170R module is installed in a PXI chasses that can directly to a computer, where it is processed by an application. The sample rate of PXIe-5170R module is $250MS/s$. The primary benefit of this DAQ system is almost dead time free, operating at a sample rate of $30000S/s$ during this experiment. We set the time window to $400ns$ and save each triggered event, while recording its trigger time at the nanosecond scale.

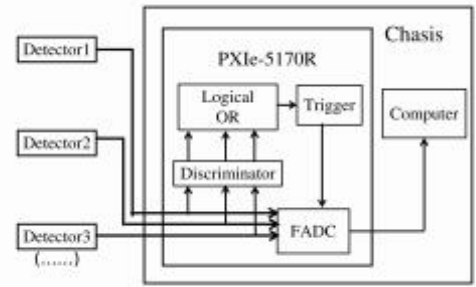

Fig. 3. Schematic diagram of the DAQ system.

### B. Experiment preparation

Prior to performing the experiment, it is essential to purge the experiment system with nitrogen. We vacuumized the experimental system to the level of $O(10Pa)$ and refilled it three times with high-purity nitrogen gas vaporized from liquid nitrogen. We used a special PTFE bottle cap equipped with a connected tee pipe to transfer the LS to the storage tank, minimizing exposure to radon. The LS bottle is connected to one end of the tee, while the other end is connected to the storage tank. The tee pipe's inter space is vacuumized before opening the valve to fill the storage tank with LS. LS is stored for at least ten times the half-life of $^{222}$Rn and thus can be used to measure radon emanations from storage tank.

### C. Experiment procedure

Since the detector made of quartz cannot be vacuumed, purged the detector and the pipeline connecting to the mixing tank with high-purity nitrogen gas at a minimum flow rate of 40L/min for 15min. This ensures the transfer of LS from the mixing tank to the detector in a high-purity nitrogen environment. We transferred 1 kg of LS from the storage tank to the mixing tank in a high-purity nitrogen environment. Then, we transferred 0.3kg of the LS from the mixing tank to a detector as a control group. When the nitrogen carrier radon has flowed through the LS for a period of time, 0.3kg of LS is transferred from the mixing tank to an identical detector. Finally, all detectors are sealed and separated from the system , then moved to a dark room surrounded by copper shields and lead bricks for further measurements.

## III. DATA ANALYSIS AND CALIBRATION

Two γ-sources , $^{137}$Cs and $^{60}$Co, were used for the energy calibration of the detector. The characteristics of the Compton edges were estimated via full GEANT4 simulations in- cluding detector resolution smearing. The spectra from the simulation and the calibration data are compared in Fig. 4. The two experimental energy points from the γ-sources as- sociated with the integral of pedestal (Ped.) were fitted with a linear function, as shown in Fig. 4.

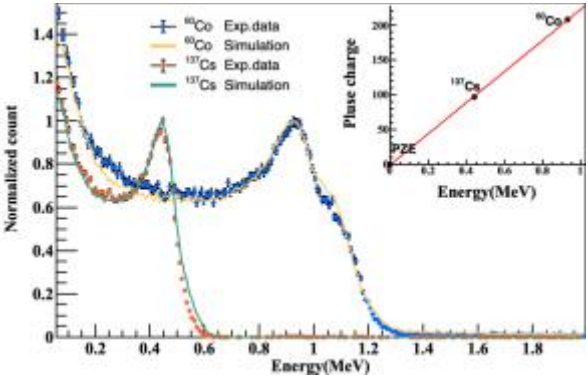

Fig. 4. (Color online) Energy calibration with $^{137}$Cs and $^{60}$Co γ-sources. The simulation results of $^{137}$Cs (green line) and $^{60}$Co (yellow line) are good agreement with the experimental data of $^{137}$Cs (orange point) and $^{60}$Co (blue point). The inset shows the calibration result using the Compton edges.

The $^{222}$Rn concentration can be measured by the decay rates of its daughter isotopes as in the following cascade decay: $^{214}$Bi-β -$^{214}$Po-α-$^{210}$Pb (Fig. 5). Pulse shape discrimination (PSD) has been utilized to distinguish nuclear recoil event(e.g. α particle) and electron recoil event as shown in Fig. 6 The discrimination factor, as defined below, is able to distinguish between beta and alpha events.

$$\text{Discrimination factor} = \frac{Q_{part}}{Q_{total}} \quad (1)$$

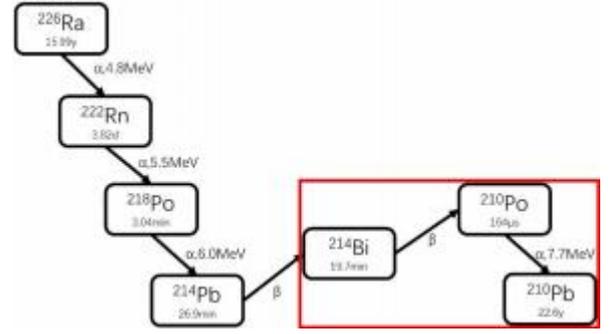

Fig. 5. Decay chain of $^{222}$Rn. Cascade decay within red box is concerned.

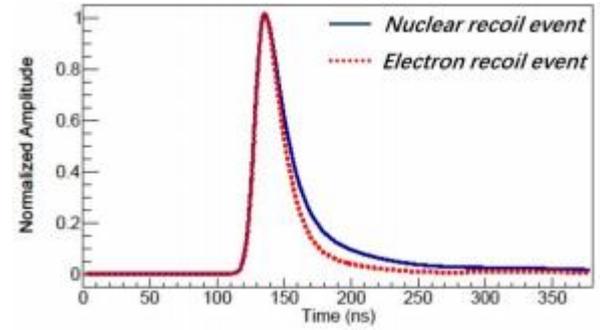

Fig. 6. Average waveform of the electron recoil event and nuclear recoil event.

$Q_{total}$ is the integral from 40ns (before the peak ) to 160ns (after the peak ), and $Q_{part}$ is the integral from 28ns (after the peak ) to 160ns (after the peak ).

As shown in Fig. 7, the LS (EJ-309) is capable of discriminating between different events.

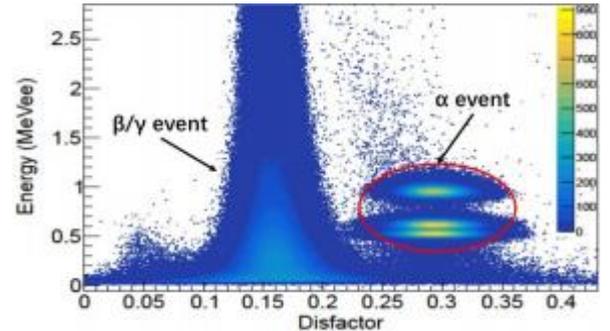

Fig. 7. Detector performance with one radon source.

To filter out the cascade decay events, three screening criteria are used: deposited energy, discrimination factor, and time interval between two events. To select the alpha events from $^{214}$Po, we found that the energy spectrum associated with alpha decay deposition is not fitted with Gaussian distribution, but with a function that combines a Gaussian and a logarithmic Gaussian, as shown in Fig. 8. Therefore, a deposit energy between 0.88–1.05 MeVee (electron equivalent) was required for the alpha event to achieve an efficiency of approximately 95%. Regarding the discrimination factor of the alpha event as shown in Fig. 9 it can be completely separated from the beta event, since the FOM (factor of merit) is approximately 4.3 and the selection between 0.22 and 0.38 corresponds to an efficiency close to 100%. The time interval between the selected event of $^{210}$Pb and the previous event of $^{214}$Po could be fitted with the exponential function to obtain the the measured half-life. The measured half-life is good agreement with the half-life of $^{214}$Po($164\mu s$), as shown in Fig. 12. In theory, the efficiency of time interval selection is 80.8% within the time window of($1/10\tau_{1/2}, 3\tau_{1/2}$). By utilizing the alpha energy selection, alpha discrimination factor selection, and the time interval selection, we analyzed the previous event (beta event) using the deposited energy and discrimination factor. As shown in Fig. 10, the discrimination factor for the beta event is almost 100% efficient within the red box. The efficiency for the beta event, where the deposited energy is greater than $0.05 \text{MeVee}$, is 96.12% compared to the simulation, as shown in Fig. 11

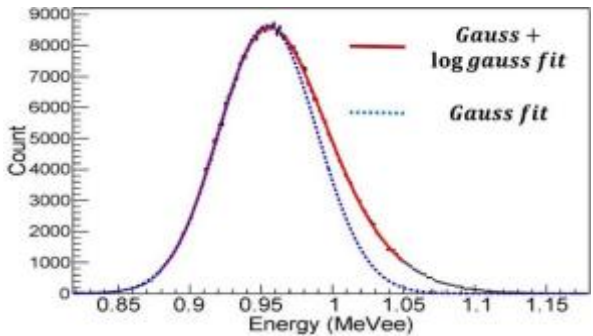

Fig. 8. Energy spectrum after filtering with discrimination factor criteria and time Interval criteria. It is found that deposited alpha energy is asymmetric.

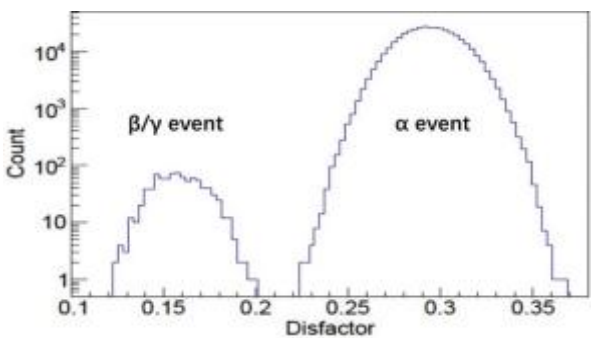

Fig. 9. Discrimination factor spectrum after filtering with energy criteria and time interval criteria. Two kinds of events can be completely separated.

The selection criteria and corresponding efficiencies for every step are summarized in the Table.1.

After subtracting the accidental background events, which amount to approximately 0.2 counts $\text{day}^{-1}$(cpd), the count rate of the selected alpha events from cascade decay of $^{214}$Bi-

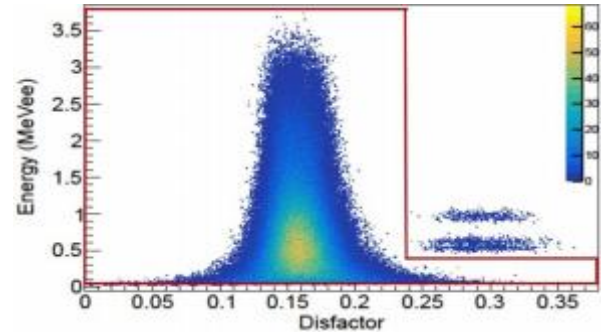

Fig. 10. All beta event candidates' distribution of energy and discrimination factor. Events within red box are selected.

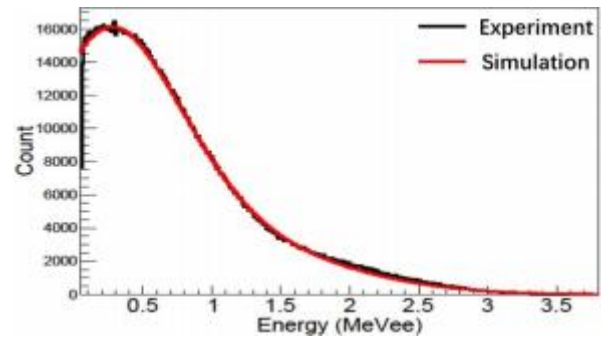

Fig. 11. Experimental deposited energy spectrum of the selected beta decay events and simulation deposited energy spectrum of the beta decay.

$^{210}$Po is plotted against realtime, as shown in Fig. 13. In addition, figure 13 shows the count rate can befitted with the Bateman equations for the decay chain of $^{222}$Rn to $^{214}$Po. The analysis concludes that $^{222}$Rn absorbed by LS reaches the transient equilibrium decay equilibrium in approximately 4h. Then we obtain the radioactivity of $^{222}$Rn at the time zero point, which corresponds to the end time point of the blowing process. Figure 14 shows that the count rate of the alpha events from $^{222}$Rn decay and $^{218}$Po decay can be plotted over time to cross-check our measurements, since the other two alpha events can be identified in the detector performance (see

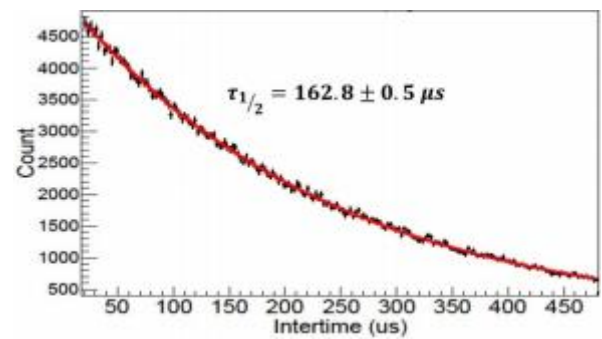

Fig. 12. Time-interval spectrum after filtering with energy criteria and discrimination factor criteria

Table 1. Selection cut and efficiency of every step.

| | Selection criteria | Efficiency |
|---|---|---|
| Alpha energy | 0.88–1.05MeVee | 95% |
| Alpha discrimination factor | 0.22-0.38 | 100% |
| Beta energy | >0.05MeVee | 96.12% |
| Beta discrimination factor | Within red box | 100% |
| Time interval | $1/10\tau_{1/2}$ -3$\tau_{1/2}$ | 80.8% |

Fig. 7).

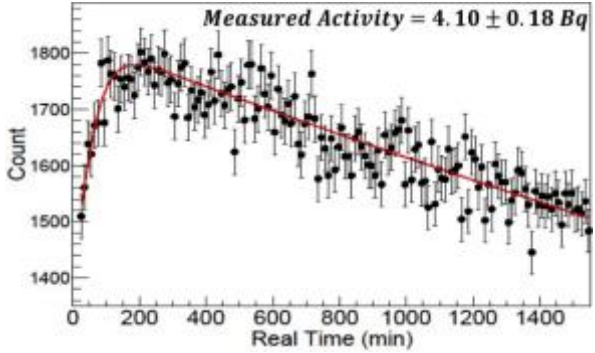

Fig. 13. Count rate of the selected alpha events from cascade decay of $^{214}$Bi-$^{210}$Po against real time. The radioactivity of $^{222}$Rn at the time zero point corresponds the end time point of the blowing process

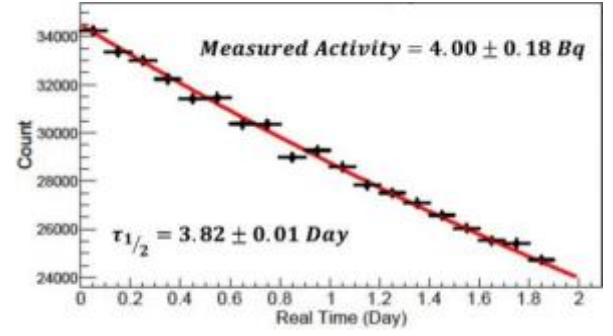

(a)

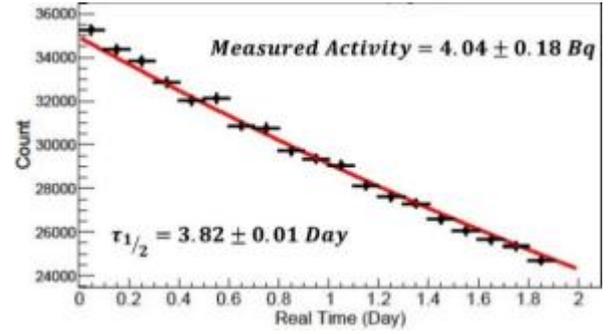

(b)

Fig. 14. Alpha count rate over realtime: (a) alpha decay of $^{222}$Rn. (b) alpha decay of $^{218}$Po.

## IV. RESULT

Prior to measuring $^{222}$Rn concentration with this system, we measured the background events of $^{222}$Rn in each section of the system.

To measure $^{222}$Rn from intrinsic $^{238}$U that originates from the detector's surface and LS contamination, the detector can be filled with LS. And the result is (30.7 ∓ 8.2)μBq/(detector). As shown in the table 2, Radioactivity of $^{222}$Rn was measured for each section of the system under specific conditions: 1.The LS was placed in the storage tank for at least ten times the half-life of Rn (seeFig. 15(b)); 2.The LS that had beenstored for an extended period in the storage tank flowed through the mixing tank (see Fig. 15(c)); 3.The LS, which had been stored in the storage tank for an extended period of time, flowed into the mixing tank, and then high- purity nitrogen gas flowed through the mixing tank at an inlet flow rate of 1L/min for 120min (see Fig. 15(d)).

Table 2. Background result of each section.

| Condition | Measured Rn radioactivity (μBq/detector) |
|---|---|
| Stored in storage tank | 28.3±14.2 |
| Flow through mixing tank | 59.8±18.4% |
| Blank group | 48.3±16.1 |

After measuring the contamination level, calibrated the coefficient of $^{222}$Rn solubility in LS using our measurement system and data analysis method. To calibrate the coefficient, we used a radon source made from $BaRa(CO_2)_3$ powder. We used high-purity nitrogen gas at a flow rate of 1L/minthrough the radon source into a commercial radon detector, RAD7, under the condition that the gas pressure is 0.1MPa, the temperature is approximately 29 °C, and the relative humidity (RH) is 2%; this was aimed at first measuring the radon source concentration and stability. The result of the source is showed in Fig. 16, and the concentration is (1600 ∓ 72)mBq/m$^3$.

Then, we use this radon source under the same condition that the flow rate is set to 1L/min, the temperature is around 29 °C, inlet gas pressure is 0.1MPa and outlet gas pressure is 0.1MPa, to flow through our experimental system for different time periods (seeFig. 17).

According to Fig. 17, Based on the data presented in Figure 17, it can be observed that the measured radon concentration in LS reaches its saturation point after approximately 69 min of blowing time.Measurement error is primarily caused by the uncertainty of radon sources. The relationship between blowing time and measured radon concentration can be modeled using Fick's law as the following diffusion function:

$$C = P_0 - P_1 \cdot e^{-t \cdot P_2} \qquad (2)$$

The results of the fitting show that parameter $P_0$ is 13.52∓0.32, parameter $P_1$ is 13.80∓0.39 and parameter $P_2$

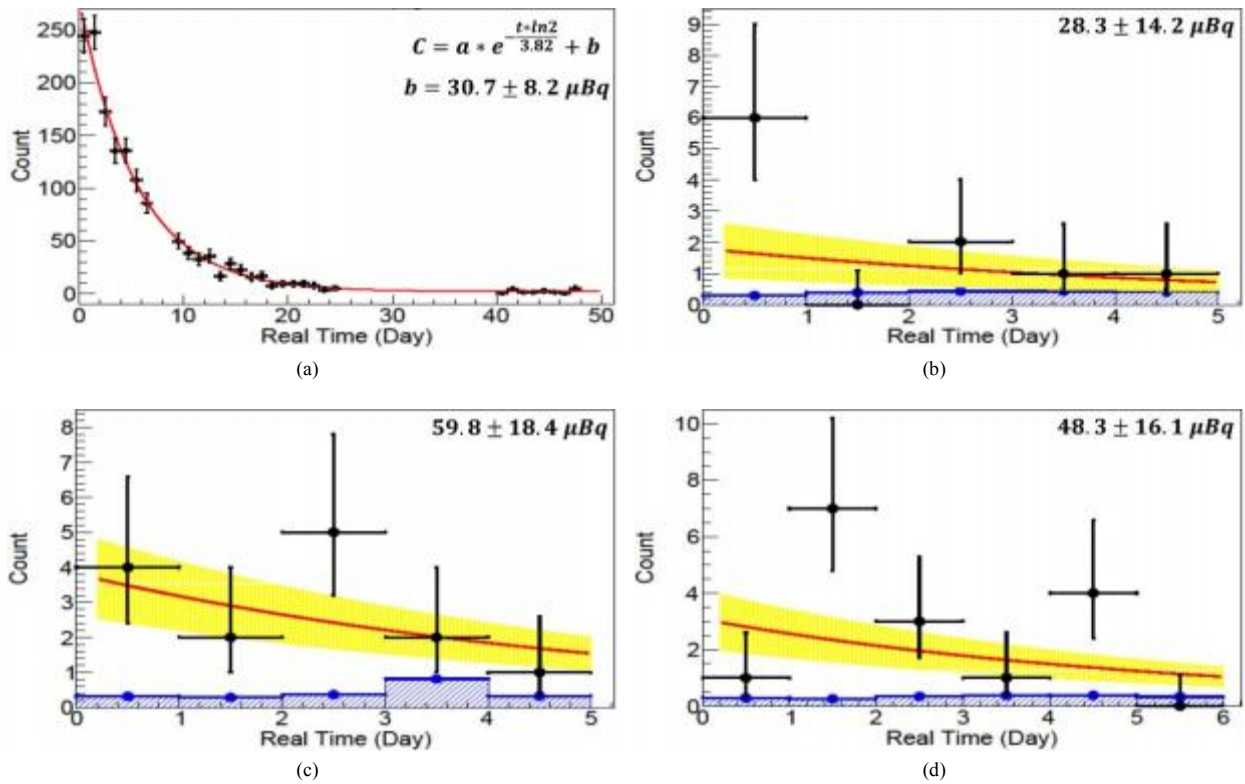

Fig. 15. Experimental results of radon background activity and fitting. Red lines represent the fitting line, black dots indicate the experimental results,and blue dots represent accidental background events. (a) detector intrinsic background; (b) storage tank background; (c) mixing tank background; (d) Blank group result.

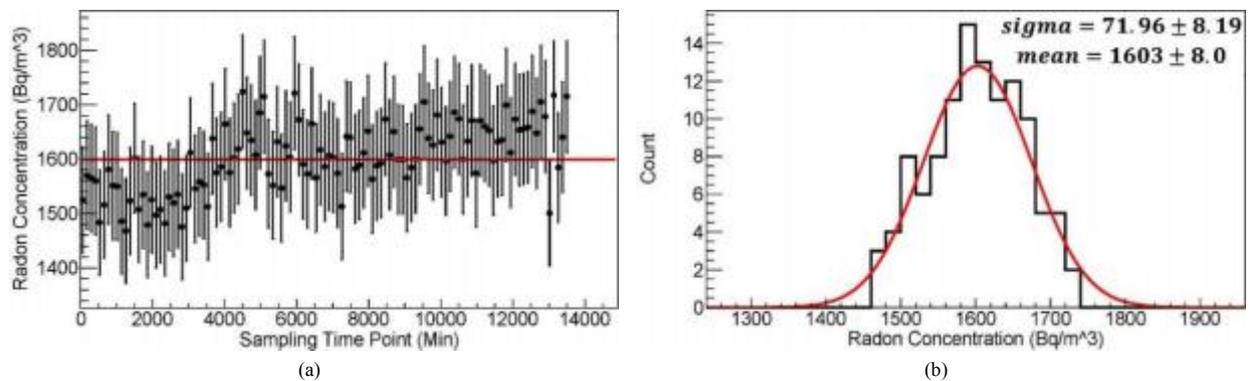

Fig. 16. RAD7 results of one certain radon source. (a) measurement fluctuation of radon concentration over time. (b) Gaussian fitting of radon concentration measurement.

is $0.067\pm0.005$. Where C is the radon concentration in LS, $P_0$ -$P_1$ 0.28Bq/kg is the saturated Rn concentration, and $P_2$ is the time factor of 99% saturation.

We conducted the same experiment with other radon sources of varying radon concentrations to establish the correlation between the radon concentrations in nitrogen and the measured saturated radon concentration in LS,as shown in Fig. 18.

The relationship between the saturated radon concentration in LS and the radon concentration in nitrogen is linear. The

absorption coefficient factor is defined as:

$$C_f = \frac{C_{LS}}{C_{N_2}} \qquad (3)$$

where $C_{LS}$ is the Rn concentration in LS, $C_{N_2}$ is the Rn concentration in nitrogen, and $C_f$ is $1.096 \times 10^{-2} \pm 5.764 \times 10^{-5}$. Therefore, at 90% confidence level, the detection limit can be defined the equation :

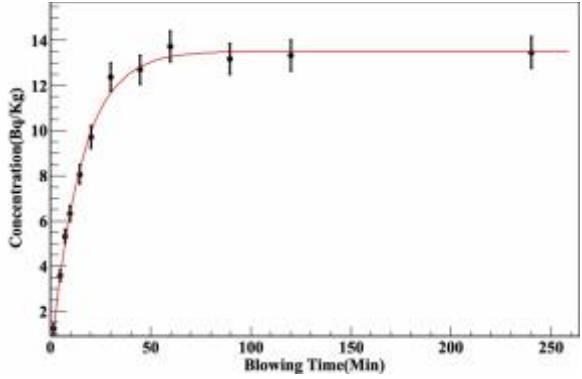

Fig. 17. Measured radon activities in LS over different blowing times.

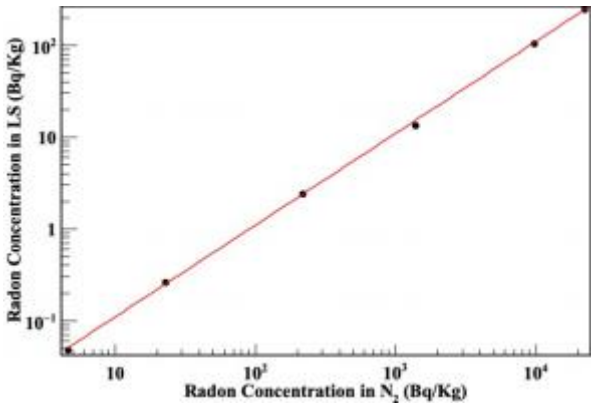

Fig. 18. Absorption co-efficiency calibration of the system. The size of dots shows the experimental error.

$$L = \frac{BG_{MAX}}{C_f} \qquad (4)$$

where L is the detection limit, $BG_{MAX}$ is the 90% confidence level of the upper limit of maximum background radon concentration, $C_f$ is the absorption coefficient factor, and the detection limit of our system is calculated to be $32 mBq/m^3$.

Furthermore, we studied the absorption coefficient factor at different temperatures and blowing pressures, as shown in the Table 3. A lower temperature, as well as a higher blowing pressure, results in a greater absorption coefficient factor. Therefore, detection limit can be optimized to be $9.6 mBq/m^3$.

## V. SUMMARY AND PROSPECT

Based on cascade decay selection, we used LS detectors to measure the radon concentration in nitrogen. Background Rn contaminations in every part of the system were measured to be less than $(59.8 \pm 18.4) \mu Bq$ per detector. Using some radon sources to carry out calibration experiments, we found that the $^{222}Rn$ absorption in LS would be eventually saturated with sufficient time and certain Rn concentrations Meanwhile, the saturated Rn concentration in LS has been found to be linearly related to the Rn concentration in nitrogen. Furthermore, the coefficient factor of Rn absorption in LS can be increased by lowering the temperature or increasing the blowing pressure; therefore, that detection limit of our system reaches $9.6 mBq/m^3$.

Table 3. Co-efficiency factor under different temperatures and blowing pressures.

| Temperature (° C) | Blowing pressure (MPa) | Flow rate (L/min) | co-efficiency factor |
|---|---|---|---|
| 29 | 0.1 | 1 | 8.68±0.25 |
| 21 | 0.1 | 1 | 9±0.26 |
| 12 | 0.1 | 1 | 9.4±0.27 |
| 21 | 0.2 | 1 | 18.8±0.54 |
| 12 | 0.3 | 1 | 28.8±0.83 |

In future studies, various aspects of the proposed measurement system can be improved. First, the volume of the LS detector can be enlarged 100 times or more to increase the accuracy of measurement significantly and relatively reduce intrinsic background events related to Rn concentration. Meanwhile, we are considering combining the storage tank, mixing tank, and detector into one tank, so that we can control the source of contamination more easily. The last and the most important goal is to develop a cryogenic charcoal trap that gathers radon to realise enrichment of radon concentration before measurement, so that the detection limit should be lowered by $O(100)$[26–29].